
\bigskip

\magnification=1200

\font\title=cmr10 scaled\magstep2
\font\small=cmr7 scaled\magstep1
\font\smallb=cmb8 scaled\magstep1

\noindent
{\title   Matrix Games, Linear Programming,  and Linear Approximation}
\bigskip

\noindent
LEONID N. VASERSTEIN, 

\noindent
{\it Department of Mathematics, Penn State U., University Park, PA 16802 }

\noindent
{\it 
(e-mail: vstein@math.psu.edu)}
\bigskip

\noindent
{\small Received  Jan. 27, 2006}
\bigskip

\noindent
{\smallb Abstract.}
{\small 
The following four classes of computational problems are equivalent:

solving matrix games, 

solving linear programs, 

best $l^{\infty}$  linear approximation,

best  $l^1$  linear approximation.}

\bigskip

\noindent
{\smallb Key words} 
{\small Matrix games,linear programming, linear approximation, least  absolute deviations.}

\bigskip
\noindent
 
\centerline{\bf Definitions}
\smallskip

First we recall relevent definitions. 

An {\it affine function}  of variables $x_1,\ldots,x_n$ is
$b_0+c_1x_1+\cdots + c_nx_n$ where $b_0,c_i$ are given numbers.

An  $l^{\infty}$ {\it  linear approximation  problem, } also known as (discrete)  {\it Chebyshev 
approximation problem }  
 is the problem of minimization
of  the following function:
$$\max(|f_1|,\ldots , |f_m|) = \|(f_1,\ldots,f_m)\|_{\infty}, \eqno{(1)}$$
where $f_1,\ldots,f_m$ are $m$ affine functions of $n$ variables.  This objective function is piece-wise linear and convex. 

An  $l^{1}$  {\it  linear approximation  problem, } also known as  
 finding the  LAD (least-absolute-deviations)    fit,
 is the problem of minimization
of  the following function:
$$\sum_{i=1}^m |f_i| = \|(f_1,\ldots,f_m)\|_1, \eqno{(2)}$$
where $f_1,\ldots,f_m$  are $m$ affine functions of $n$ variables.  This objective function is piece-wise linear and convex.

A {\it matrix game} is given by a (payoff) matrix  $A.$ To solve a matrix game is
to find a row $p$ (an optimal strategy for the row player), a column $q$
(an optimal strategy for the column player), and a number $v$ such that
$p =(p_i) \ge 0, \sum p_i = 1, q =(q_j) \ge 0, \sum q_i = 1, 
 pA \ge v \ge Aq.$ The number $v$ is known as the value of game.
The pair $(p,q)$ is   known as an equilibrium for the matrix game.

As usual, $x \ge 0$ means that every entry of the vector $x$ is  $\ge 0.$
We write $y \le t$ for a vector $ y $ and a number $t$ if every entry of $y$ is $\le t.$  
We go even further in abusing notation, denoting
by $y-t$ the vector obtaining from $y$ by subtracting $t$ from every entry.
Similarly we denote by $M+c$ the matrix obtained from $M$ by adding a number $c$ to every entry.

A matrix game is called {\it symmetric}  if the payoff matrix is skew-symmetric.
Recall that the value of any symmetric game is 0, and the transposition gives  a bijection between the optimal strategies of the players.

A {\it linear constraint}  is any of the following constraints: $f \le g, 
f \ge g,  f = g, $ where $f,g$ are affine functions.
A {\it linear program} is an optimization (maximization or minimization) of an affine function
subject to a finite system of linear constraints.

\medskip

\centerline{\bf Statement of results}
\smallskip

It is well known, that solving a matrix game can be reduced to solving a pair of linear programs, dual to each other. It is also known that solving any linear program
can be reduced to finding an optimal strategy with positive last component
for  a symmetric matrix game.  In both reductions, the size of data (in terms of the number of given
numbers or the number of given bits)  may increase at most two times.

A subtle point here is: how can we compute an optimal strategy (for a symmetric game)  with  a positive last entry or prove that no such strategy exists?   An answer is that for any vertex in the set of optimal strategy with positive last entry
is a solution of a system of linear equations   whose coefficients are the entries of the payoff matrix or 0,1, so a positive lower bound  $\alpha$ can be given for this entry (at least in the case when all given
numbers are rational).   Namely, let $\beta$  be an upper bound for the absolute values of the numerators
and denominators of the entries of the payoff matrix of size $N$ by $N. $ Then $\alpha = \beta^{-2N}N^{-N/2}$ will work. Notice that   $0 < \alpha <  1.$

The mixed strategies for the column player
with the last entry $\ge \alpha$
 in the symmetric game
 are the mixed strategies    for the column player for the modified game obtained by adding the 
 $(\alpha/(1-\alpha))$-multiple of the last column
 to the other columns of the payoff matrix.
The optimal strategies for a modified matrix game give
optimal strategies with positive last entry for the original symmetric game
provided that the value of the modified game stays 0 (otherwise, there are no optimal strategies with positive last entry for the 
original symmetric game hence the original linear program has no optimal solutions).

Given any $\l^{\infty}$  approximation problem with the objective function  (1),  here is a well-known reduction (Vaserstein, 2003)  to a linear program
with one additional variable $t$:
$$t \to \min, \  {\rm subject\  to} \  -t \le f_i  \le t \  {\rm for } \  i =1,\ldots, m.$$
This is a linear program with $n+1$ variables and $2m$ linear constraints.
Since any linear program can be reduced to a matrix game (see above), we conclude that finding
an Chebyshev fit can be reduced to solving a matrix game. 

 The converse reduction is a main goal of this paper:

{\bf Theorem 1}.  Solving any matrix game  can be reduced to finding a Chebyshev fit.
More precisely, when the game is given  by an $m$ by $n$ matrix,
we construct  a Chebyshev  approximation problem  with  $2m+2n+3$ affine functions of $m+n+1$ variables as well as  a bijection
between the equilibria for the matrix game and the solutions for the   approximation problem.

Given any $l^1$  approximation problem with the objective function (2),  here is a well-known reduction (Vaserstein, 2003)  to a linear program
with $m$  additional variables $t_i$:
$$\sum_{i=1}^m t_i \to \min, \  {\rm subject\  to} \  -t_i \le f_i  \le t _i\  {\rm for } \  i =1,\ldots, m.  $$
This is a linear program with $n+m$ variables and $2m$ linear constraints.
Since any linear program can be reduced to a matrix game (see above), we conclude that finding
the best $l^1$-fit can be reduced to solving a matrix game. 

 The converse reduction is  the second goal of this paper:
 
 {\bf Theorem 2}.  Solving any matrix game  can be reduced to solving
 an $l^1$ linear approximation problem.
More precisely, when the game is given  by an $m$ by $n$ matrix,
we construct  an $l^1$ approximation problem  with $4m+4n+6$   affine functions of $m+n+1$ variables as well as  a bijection
between the equilibria for the matrix game and the solutions for the   approximation problem.

\medskip
\centerline{\bf Proof of Theorem 1}
\smallskip

Consider   any matrix game with the payoff matrix  $A$  with $m$ rows and $n$ columns.
It can  can be reduced to 
the symmetric game  with  the payoff matrix  
$$M= \pmatrix{0 & A+C  & -J \cr -A^T-C &  0 &  J' \cr J^T & -J'  &0},$$
where $J$ (rest. $J'$) is the column of $m$ (resp., $n$) ones and
the number $C$ is   such that    $ A+C > 0.$ 
The skew-symmetrix matrix $M = -M^T$ has size $(m+n+1) \times (m+n+1).$
(J. von Neumann suggested another reduction resulting in a skew-symmetric matrix of size $(mn) \times (mn)$ which is not so good from computational point of view.)

The bijection between the  solutions  $(p,q,v)$ for the game with the matrix $A$ and the optimal strategies for the row player in the symmetric game with the matrix $M$ is given by

$$(p,q)  \mapsto (p,q^T,v+C)/(2+v+C).$$

Note that the last entry of any optimal strategy for the symmetric game above is positive because $A+C > 0.$

Now we start with any matrix game, with the payoff matrix $M = - M^T$ of size $N$ by $N.$
(In the situation above, $N = m+n+1.)$ Our problem is to find   a column $x=(x_i)$ (an optimal strategy)
such that  
$$  Mx \le 0, x \ge 0, \sum x_i = 1. \eqno{(3)}$$

This problem (3) (of finding an optimal strategy)  is about finding a feasible solution for a system of linear constraints. It can be written as the following linear program with an additional variable $t$ and the optimal value 0:
$$  t \to \min, Mx \le t,  x \ge 0, \sum x_i = 1.\eqno{(4)}$$

Now we find the largest entry $c$ in the matrix $M$.  If $c=0,$
then $M=0$ and the problem (1) is trivial (every mixed strategy $x$ is optimal). So we assume that $c > 0.$

Adding the number $c$  to every entry of the matrix
$M,$ we obtain a matrix $M+c \ge 0 $ (all entries $\ge 0).$  The linear program (4)
is equivalent to
$$  t \to \min, (M+c)x \le t,  x \ge 0, \sum x_i = 1 \eqno{(5)}$$
in the sense that these two programs have the same feasible solutions and  the same optimal solutions. The optimal value for (4) is 0 while the optimal value for (5) is $c.$

Now we can rewrite (5) as follows:
$$  \|(M+c)x \|_{\infty}  \to \min,  x \ge 0, \sum x_i = 1 \eqno{(6)}$$
which is a Chebyshev approximation problem with additional linear constraints.
We used that $M+c \ge 0,$ hence  $(M+c)x  \ge 0$ for every feasible solution $x$ in (4).
The optimal value is still $c.$

Now we rid off the constraints in (4)  as follows:
$$ \|\pmatrix{(M+c)x \cr c-x \cr  \sum x_i +c - 1 \cr -\sum x_i -c + 1}  \|_{\infty} \to \min. \eqno{(7)}$$

Note that the optimization problems (6) and (7) have the same optimal value
$c$ and every optimal solution  of (6) is optimal for (7).
Conversely, 
 for every $x$ with a negative entry, the objective function in (7) is
$> c.$ Also, for  every $x$ with  $\sum x_i \ne 1,$  the objective function in (7) is
$> c.$ So every optimal solution  for (5) is feasible and hence optimal for (6).

Thus, we have reduced solving any symmetric matrix game with $N \times N$ payoff matrix to a Chebyshev approximation problem (7) with $2N+2$ affine functions in $N$ variables.

\medskip
\centerline{\bf Proof of Theorem 2}
\smallskip

As in the proof of Theorem 1,  we first reduce our game to a symmetric $N$ by $N$ game where  $N = m+n+1$  and set $c$ to be  largest entry in the matrix $M.$ The case $c=0$ is trivial, so let $c > 0.$

We want to find a column $x$ such that 
$$
x \ge 0, \sum x_i = 1, Mx \le 0.
$$
Consider the   $l^1$  approximation problem whose objective function is $f(x)=$
$$\| \pmatrix{Mx \cr c+Mx  \cr x \cr 1-x \cr -1+ \sum x_i  \cr 1 -  \sum x_i} \|_1 =
\| Mx \|_1 +\| c+Mx \|_1 +
\| x \|_1 + \|1-x\| _1+ \|-1+ \sum x_i \|_1 +\| 1 - \sum x_i \|_1
$$
with $4N+2$ affine functions of $N$ variables.

Note that  $f(x) =Nc+N$ for every optimal strategy $x$ and that
$f(x) > Nc+N$ for every   $x$ which is not an optimal strategy.
So solving this approximation problem is equivalent to solving the matrix game.
\smallskip

{\bf  Remark. }  
 Our result implies that every  $l^{1}$ linear approximation problem can be reduced to a
 $l^{\infty}$ linear approximation problem and vice versa..
 
There is 
 an obvious    direct reduction  of the $l^1$ approximation problem with the objective function  (2) to
 $$\max| f_1\pm f_2 \pm \cdots \pm f_m| \to \min
 $$
 which is a Chebyshev approximation problem  with  $2^{m-1}$ affine functions in $n$ variables.
 This reduction increases the size exponentially, while our reductions increases the size linearly.
 
 \smallskip
 
 {\bf  Remark. }   There are methods for solving $l^1$ approximation problems
 alternative to the simplex method  [Bloomfield--Steiger 1983]. Our reductions allows us to use these methods for solving arbitrary linear programs and matrix games.
  \smallskip
 
 {\bf  Remark. }   A preprint with Theorem 1 appeared at arXiv
 [Vaserstein 2006].

 \bigskip
 \noindent
 \centerline{\bf References}
 \smallskip
 
 \noindent
 Bloomfield, Peter and   Steiger, William L. {\it 
Least absolute deviations.
Theory, applications, and algorithms.}    Progress in Probability and Statistics, 6.
BirkhŠuser Boston, Inc., Boston, MA, 1983. xiv+349 pp. ISBN 0-8176-3157-7.
 
 \smallskip
 \noindent
Vaserstein, L. N. (2003),  {\it  Introduction to Linear Programming,}  Prentice Hall. 
(There is a Chinese translation by    Mechanical Industry Publishing House ISBN: 7111173295.)

 \smallskip
 \noindent
 Vaserstein L.N.,  
  Reduction of Linear Programming to Linear Approximation,
   arXiv.org,   math.OC/0602339,  15 Feb 2006.

\end